\documentclass[referee]{aa}
\usepackage{graphics}

\def\xibar{\bar{\xi}}
\def\gammastar{\Gamma^*}
\def\h{$h^{-1}$ Mpc}

\begin{document}

\thesaurus{2(12.03.3; 12.12.1)}

\title{Is the Universe a fractal? Results from the SSRS2}

\author{A. Cappi\inst{1}, C. Benoist\inst{2}, L.N. da Costa\inst{2,3}, 
        S. Maurogordato\inst{4}}
\institute{Osservatorio Astronomico di Bologna, via Zamboni 33, I-40126,
   Bologna, Italy
\and
European Southern Observatory, Karl-Schwarzschild-Str.2,
   D-85748 Garching bei M\"unchen, Germany
\and
Departamento de Astronomia CNPq/Observat\'orio Nacional, 
rua General Jos\'{e} Cristino 77, Rio de Janeiro, R.J. 20921 Brazil
\and
CNRS, UMR 6527; 
Observatoire de la C\^ote d'Azur, B4229, Le Mont--Gros, F-06304 
Nice Cedex 4, France
}
\offprints{Alberto Cappi}
\mail{cappi@astbo3.bo.astro.it}
\date{Received ~ / Accepted ~}
\titlerunning{Is the Universe a fractal?}
\authorrunning{A. Cappi et al.}
\maketitle
\begin{abstract}
We perform a fractal analysis of the Southern
Sky Redshift Survey 2, following the
methods prescribed by Pietronero and collaborators, to check
their claim that the galaxy distribution is fractal at all scales, and
we discuss and explain the reasons of some controversial points, 
through tests on both galaxy samples and simulations.
We confirm that the amplitude of the two--point correlation function does not 
depend on the sample depth, but increases with luminosity.
We find that there is no contradiction between the results of
standard and non--standard statistical analysis;
moreover, such results are consistent with theoretical
predictions derived from standard CDM models of galaxy formation,
and with the hypothesis of homogeneity at large scale
($\sim 100$ \h).
However, for our SSRS2 volume--limited subsamples we show that
the first zero--point of the autocorrelation function $\xi(s)$
increases linearly with the sample depth,  and that its value is
comparable to the radius of the maximum sphere which
can be completely included in the sampled volume;
this implies that the true zero--crossing point of $\xi(r)$ has not been 
reached. We conclude that the apparent fractal behavior is due to 
a combination of a luminosity--dependent
correlation amplitude and the recovering of power at larger scales
in deeper samples.

\keywords{cosmology: observations --- large-scale structure of universe}
\end{abstract}

%
%
\section{Introduction}

One of the pillars of the standard cosmological models is the
large--scale homogeneity of the Universe (e.g. Peebles 1993).
The standard statistical methods to analyze the large--scale structure,
as the two--point correlation function $\xi(r)$, 
are described by Peebles (1980); they rely
on the definition of a mean galaxy density $\bar{n}$, 
which is meaningful only if the assumption of large--scale
homogeneity is true.
However, at small scales the autocorrelation function of the galaxy 
distribution is positive and can be fitted by a power--law, 
which is a property of a fractal set (see Peebles 1980). On the basis
of the observational evidence that galaxies are clustered in ever
increasing systems, from groups and clusters to superclusters, 
de Vaucouleurs (1970, 1971) presented
``the case for a hierarchical cosmology''\footnote{The idea of a
 hierarchical Universe has a long history, which dates back to the
XVIII$^{th}$ century (see  E. Harrison, 1987, 
{\em Darkness at night, A Riddle of the Universe},
Harvard University Press, Cambridge); in this century,
Fournier d'Albe and Charlier were the first
to propose a hierarchical (now we would say fractal) model of the
Universe (see Mandelbrot 1982).}.
Mandelbrot (1982) developed this concept,
suggesting that the galaxy distribution in the Universe
is fractal, with dimension $D = 1$.

In addition, in the last 20 years redshift surveys at
increasing depths have revealed ever larger structures and voids
(see e.g. Davis et al. 1982; de Lapparent et al. 1986, da Costa et al. 1994,
Vettolani et al. 1997, and references therein).
Einasto et al. (1986) found evidence that $r_0$ increases with sample volume,
but they estimated that it should reach a value $\sim 10$\h~ for
a fair sample of the Universe. 
Pietronero (1987) stressed that, if homogeneity is not reached, 
the correlation length $r_0$
cannot be taken as a measure of the clustering amplitude,
and suggested a slightly but significantly different definition
of the autocorrelation function. Adopting this approach,
Coleman et al. (1988) reanalyzed the CfA1 redshift 
survey (Huchra et al. 1983), concluding that the distribution of 
galaxies was fractal to at least 20 \h. On the other hand,
Davis et al. (1988) found that $r_0$ increased as $r_0^{0.5}$, and
not linearly with the depth as predicted for a simple fractal 
(see also Maurogordato et al. 1992).
Others advocated the need for a multifractal approach 
(e.g. Balian \& Schaeffer 1989; Martinez \& Jones 1990; 
Martinez et al. 1990); this is however another issue and 
we will not discuss it: see for example the review by Borgani (1995) and 
references therein.

The first redshift surveys sampled relatively small
volumes, and the reality and nature
of correlation amplitude variations with depth remained
an open --and much debated-- question. Could these variations
simply reflect fluctuations due to local structures? Were they
a consequence of the fractal distribution of galaxies?
Or were they an indirect consequence of the dependence of
clustering on galaxy luminosity? 

It should be pointed out that even at a scale $\sim 1000$ \h~ 
we do not expect a {\em perfect} homogeneity, as COBE has
found evidence of anisotropy in the CMB radiation at a level 
$\delta T / T \sim 10^{-5}$ (Smoot et al. 1992), 
a value necessary and sufficient, at least in some standard cosmological
models, to explain the formation of the observed structures. 
The existence of very large structures in the Universe implies that the
galaxy or cluster autocorrelation function must be positive to a scale
corresponding to the size of these structures; but 
this is not necessarily inconsistent with the measured
value of the galaxy autocorrelation length $r_0 \sim 5-6$ \h~ 
for $L \sim L_*$ galaxies (as claimed by Pietronero and
collaborators), as larger structures have also a lower contrast,
and the correlation of luminous matter may be significantly
amplified relatively to the underlying dark matter correlation function
(Kaiser 1984; Bardeen et al. 1986).
Therefore, we expect that the galaxy and cluster distribution will not be 
perfectly homogeneous even at 
large scales. The claim for a fractal Universe is obviously much more 
stronger than that: it implies that there is no convergence to 
homogeneity and that it is not 
possible to define a mean density $\bar{n}$ of the Universe.

In the last years Pietronero and his collaborators
applied statistical indicators which do not imply a universal mean density 
on an ever increasing number of catalogs 
(see e.g. Di Nella et al. 1996, Sylos Labini,
Montuori \& Pietronero 1996, Montuori et al. 1997), 
claiming evidence for a galaxy fractal 
distribution at all scales. Their results are impressive, as 
recently stressed by Coles (1998), but they appear to be
at variance with the results
of other groups who analyzed the same catalogs with standard indicators.
It is clear that the situation is unsatisfactory,
despite many articles and even
public debates on the subject (see Pietronero et al. 1997; Davis 1997;
Guzzo 1997); while the strongest support to large--scale homogeneity
comes indirectly from the high level of isotropy of the CMB radiation and
from two--dimensional catalogs (Peebles 1996), there is still confusion on the 
interpretation of the available three--dimensional data, mainly due
to the difference in the statistical indicators, which do not allow
a direct and quantitative comparison of the results.

Therefore we analyzed the Southern Sky Redshift Survey 2 
(SSRS2; da Costa et al. 1994) following the approach of Pietronero and
collaborators, in order to independently check their claims
and to answer to their criticisms (Sylos Labini et al. 1997, hereafter SMP) 
about the work of Benoist et al. (1996, hereafter Paper I).
In section 2 we discuss the apparent dependence of $r_0$ on sample depth;
in section 3 we describe the different statistics, and their relations;
in section 4 we present and discuss our fractal analysis of the SSRS2 
in comparison with the standard analysis;
in section 5 we show that theoretical predictions derived from the
standard CDM model can reproduce our results, and
are therefore consistent with a homogeneous Universe;
our conclusions are in section 6. 

%
%
\section{Does $r_0$ depend on sample depth?}

If the Universe is homogeneous at large scales, then for any
fair sample the statistical properties of the galaxy distribution
should be the same; in particular, the autocorrelation function of any
given class of galaxies should not depend on the sample depth.
On the contrary, if the Universe is a simple fractal at all scales, 
the mean density in a sphere of radius $r$ centered on a randomly chosen 
galaxy will vary as 

\begin{equation}
\label{eq:nfrac}
 n(r) \propto r^{D-3}
\label{eq:fracdens}
\end{equation}

where $D$ is the so--called correlation dimension.

The autocorrelation function will vary as

\begin{equation}
\label{eq:xifrac}
\xi(r) \propto r^{(3 - D)} - 1
\end{equation}

and the correlation length $r_0$ will
increase with the sample depth:

\begin{equation}
r_0 = [(3 - \gamma)/6]^{1/\gamma} R_s
\end{equation}

where $\gamma = 3 - D$ and $R_s$ is a measure of the sample size.

\begin{figure}
\resizebox{\hsize}{!}{\includegraphics{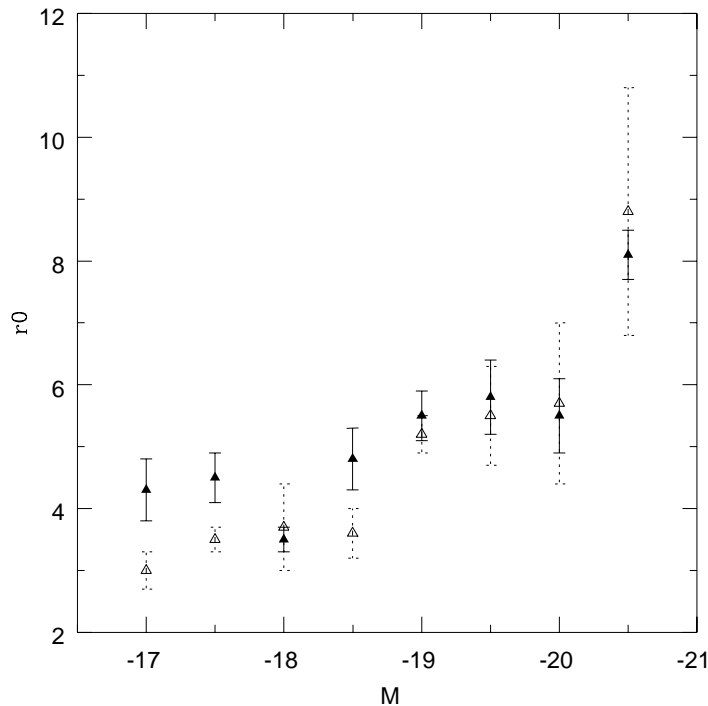}}
\caption{Correlation length $r_0$ for different volume--limited subsamples
extracted from the SSRS2, assuming an apparent limiting
magnitude $m_{lim} = 14.5$ (open triangles) and $m_{lim} =15.5$ 
(solid triangles)}
\label{fig:depth}
\end{figure}

In practice, there is the problem to define a sample not dominated by 
a few prominent structures, and to discriminate between a dependence
of $r_0$ on the sample depth or on the absolute luminosities of the galaxies, 
as deeper volume--limited samples extracted from a magnitude--limited catalog 
will progressively include more luminous galaxies.
The most recent results appear to confirm the original claims
by Hamilton (1988) and B\"orner et al. (1989) about a real dependence
of $r_0$ on the luminosity. In particular, in Paper I we analyzed the
SSRS2 and showed that $r_0$ has a significant increase 
for $L \ge L_*$ galaxies. SMP criticize these results and
point out that a value of $r_0$ as large as 16 \h, as found
in our Paper I for the $M \le -21$ galaxies, has never been
found in previous surveys as the CfA1.
We stress however that 
a) the previous, shallower surveys do not contain enough 
galaxies with $M \le -21$ galaxies which, being rare, require a larger
volume; b) as shown in figure 5 of Paper I, the
results obtained by Hamilton on the CfA1 are consistent 
with the new SSRS2 results
--up to  $M=-20.5$, representing the maximum
luminosity reachable in a volume--limited
sample large enough for a statistical
analysis in the CfA1--.
Notice that the CfA1 and the SSRS2 are limited respectively to $m = 14.5$ and 
$m = 15.5$, so volume--limited samples containing galaxies with the same 
absolute luminosity have different depths; their consistency at a fixed
absolute magnitude is another argument in favor of luminosity segregation
and against a fractal distribution. However, SMP criticize
our Paper I also because we tested the independence of 
the correlation length from the sample depth
only for a few subsamples; moreover, they notice that the results
obtained by Sylos Labini \& Pietronero (1996) on the Perseus--Pisces 
catalog show a systematic dependence of  $r_0$ on the sample depth,
and not on luminosity.
Indeed the same test performed by SMP on the Perseus--Pisces is
possible also for all the subsamples of the SSRS2. 
In figure (\ref{fig:depth}) we present $r_0$ with its bootstrap error
as a function of the absolute magnitude for volume--limited samples of
the SSRS2 selected with apparent limiting magnitudes of 14.5 and 15.5 
(see table 1 and 2). Notice that the distance limits for the volume--limited
samples are calculated assuming $H_0 = 100$~km~s$^{-1}$~Mpc$^{-1}$ and
taking into account the K--correction, and they are conservatively fixed to
include the galaxies of all the morphological types (for more
details see Paper I).
%
%
\begin{table}
\caption[]{Volume--limited subsamples from the SSRS2}
\begin{flushleft}
\begin{tabular}{rrrrr}
\hline
 $M_{lim}$ & $D_{L}$ & $D_{com}$ & z$_{lim}$ & $N_{gal}$ \\
\hline
 -16.0 &  19.70 &  19.58 & 0.0066 & 284 \\
 -16.5 &  24.73 &  24.53 & 0.0082 & 279 \\
 -17.0 &  31.01 &  30.69 & 0.0103 & 302 \\
 -17.0 &  38.84 &  38.35 & 0.0129 & 329 \\
 -18.0 &  48.60 &  47.83 & 0.0161 & 488 \\
 -18.5 &  60.72 &  59.52 & 0.0202 & 773 \\
 -19.0 &  75.72 &  73.87 & 0.0251 & 766 \\
 -19.5 &  94.23 &  91.38 & 0.0312 & 755 \\
 -20.0 & 116.95 & 112.60 & 0.0386 & 593 \\
 -20.5 & 144.71 & 138.12 & 0.0477 & 268 \\
 -21.0 & 178.41 & 168.52 & 0.0587 & 133 \\
\hline
\end{tabular}
\end{flushleft}
\end{table}
%
%
\begin{table}
\caption[]{Volume--limited subsamples from the SSRS2 with a limiting
  apparent magnitude 14.5}
\begin{flushleft}
\begin{tabular}{rrrrr}
\hline
 $M_{lim}$ & $D_{L}$ & $D_{com}$ & z$_{lim}$ & $N_{gal}$ \\
\hline
-16.5 &  ~15.69 &  ~15.61 & 0.0053 & ~97 \\
-17.0 &  ~19.70 &  ~19.58 & 0.0066 & 164 \\
-17.5 &  ~24.73 &  ~24.53 & 0.0082 & 156 \\
-18.0 &  ~31.00 &  ~30.69 & 0.0103 & 155 \\
-18.5 &  ~38.84 &  ~38.34 & 0.0129 & 141 \\ 
-19.0 &  ~48.60 &  ~47.83 & 0.0161 & 185 \\
-19.5 &  ~60.72 &  ~59.51 & 0.0215 & 212 \\
-20.0 &  ~75.72 &  ~73.87 & 0.0251 & 134 \\
-20.5 &  ~94.23 &  ~91.38 & 0.0312 & ~68 \\
\hline
\end{tabular}
\end{flushleft}
\end{table}
%
%

As discussed in Paper I, the less deep subsamples (including galaxies
with a faint absolute magnitude) may be affected by the dominance of local 
structures. From figure \ref{fig:depth},
there is indeed marginal evidence for sampling variations in
subsamples including galaxies with absolute magnitude fainter than 
$M = -19$, while for deeper subsamples 
galaxies with the same absolute magnitude have the same value of $r_0$
within the errors, independently of the sample depth.
This consistency internal to the SSRS2, in addition
to the external consistency between subsamples extracted from the CfA1
and the SSRS2, is a further and robust
confirmation that the correlation amplitude depends on the luminosity
and not on the sample depth. This result is obviously
in disagreement with the claims by
SMP, based on subsamples of the Perseus--Pisces supercluster, which
show indeed very large differences between the
correlation length $r_0$ of galaxies with the same absolute magnitude 
but in volumes at different depths (see their figure 40). 
SMP assume a constant error $\Delta r_0 = 0.5$ \h; while a more
realistic error would have been preferable, it cannot be a large 
underestimate given the size of their subsamples.

While there is no {\em a priori} reason
to prefer one catalog to another, the type of information which
can be extracted depends on the particular selection criteria of each 
catalog. In this case, contrarily to the ``anonymous'' CfA or SSRS 
redshift surveys, the name Perseus--Pisces redshift survey reminds us
that it is a region of the sky selected because of the presence,
at about $5500$ km/s, of a large filamentary structure 
(Giovanelli \& Haynes 1991).
While many interesting statistical results can --and could-- be obtained,
their interpretation must take into account this peculiarity
of the survey. For example, Guzzo (1997) discusses the effect of the 
Perseus--Pisces supercluster on the galaxy counts, questioning the 
interpretation of Pietronero et al. (1997).
As concerning the variation of $r_0$ with sample depth, 
we notice that a volume--limited sample including all galaxies 
brighter than -19 has a depth of 
$\sim 5000$ km/s at an apparent limiting magnitude of $m = 14.5$, 
and a depth of $\sim 8000$ km/s 
at an apparent limiting magnitude of $m = 15.5$.
In the case of the Perseus--Pisces redshift survey,
this means that in the first case we include the 
void in front of the supercluster but not the supercluster itself, while 
in the second case we include the whole supercluster.
This might at least partly explain the discrepancy with our results; 
therefore a fractal galaxy distribution does not appear
as the most plausible explanation for the systematic variations of
$r_0$ in the Perseus--Pisces redshift survey.

%
%
\section{Statistical tools}

As we mentioned in the introduction, the 
statistical indicators used to assess the fractality of the Universe
are different from --but not independent of-- the standard ones.
In this section we will clarify the relations between the two sets of
indicators.

A simple dimensionless measure of galaxy clustering is the two--point
correlation function $\xi(\vec{r})$, which is defined by the joint
probability $\delta P$ of finding a galaxy in a volume element $dV_1$
and another galaxy in a volume element $dV_2$ at separation $\vec{r}$:
$\delta P = \bar{n}^2 [1 + \xi(r)] dV_1 dV_2$,  
where we take into account only the magnitude of the separation but not
its direction, assuming as usual that the galaxy distribution is a stationary 
point--process (see Peebles 1980).
The two--point correlation function $\xi(r)$ of a sample
is measured generating a catalog of randomly distributed points in the same 
volume, with the same selection function, 
then counting the number $N_{gg}(r)$ of distinct 
galaxy--galaxy pairs at separation $r$, and the number $N_{gr}(r)$ of 
galaxy--random pairs at separation $r$ (Davis \& Peebles 1983):

\begin{equation}
\xi(r) = 2 \frac{n_R}{n_G} \frac{N_{gg}(r)}{N_{gr}(r)} - 1
\label{eq:nxi}
\end{equation}

where $n_G$ is the mean galaxy density and $n_R$ the density of points
in the random catalog.
Hamilton (1994) suggested a slightly different estimator of $\xi(r)$
less dependent on the density, and best suited to measure $\xi(r)$
at large separations, which we adopted in Paper I, and which we
use also in this paper. This is what we will call in the following
the ``direct'' estimate of $\xi(r)$.

The galaxy two--point correlation function is well approximated by
a power--law, $\xi(r) = (r/r_0) ^{- \gamma}$, 
where $\gamma \sim 1.8$, at least at scales less than $\sim 10$\h:
this implies that the correlation dimension $D = 3 - \gamma$ is $\sim 1.2$.

As discussed in the previous section,
if the Universe is a fractal, it is impossible to find a
``universal'' value of $r_0$, 
because larger samples will have larger voids and a larger $r_0$.
For this reason Pietronero (1987) has defined the conditional average
density $\Gamma(r)$:

\begin{equation}
\Gamma(r) = \frac{1}{N} \sum _{i=1} ^N \frac{1}{4 \pi r^2 dr}
\int _r ^{r + dr} n(r_i + r) dr
\end{equation}

which measures the average density at distance $r$ from an occupied point.
$\Gamma(r)$ is in principle related to $\xi(r)$:

\begin{equation}
\xi(r) = \Gamma(r) / n - 1
\label{eq:xigam}
\end{equation}

One implication  (trivial but important) of relation (\ref{eq:xigam}) 
is that, if $\xi(r)$ is a power--law, $\Gamma(r)$ is not, and vice--versa. 
Moreover, there is a practical difference in the way of estimating the
two functions (see e.g. Coleman \& Pietronero 1992).
$\Gamma(r)$ is calculated only within the sample limits, while
$\xi(r)$, as apparent from equation (\ref{eq:nxi}), is computed
using random catalogs to correct for volume boundaries, 
thus implicitly assuming homogeneity. 
Even if the Universe is not a fractal, this
can be a problem when the sample size is still small relatively to the scale 
where homogeneity is achieved;
as a consequence, an indirect estimate of $\Gamma(r)$ obtained 
from $\xi(r)$ through equation (\ref{eq:xigam}) may be wrong.

$\xi(r)$ is however a simple statistical measure of clustering; 
other useful functions representing variously defined
integrals of $\xi(r)$ can be directly measured on a sample.

For example, the mean number of objects $\bar{N}_p(r)$ in a
volume $V$ centered on a randomly chosen object, is

\begin{equation}
\bar{N}_p(r) = \int _0 ^r [1 + \xi(x)] dV = 4 \pi \int _0 ^r x^2 [1 + \xi(x)]
dx = nV + 4 \pi J_3
\label{eq:ncen}
\end{equation}

where $J_3(r)$ is the so--called $J_3$ integral: 

\begin{equation}
J_3(r) = \int _0 ^r x^2 \xi(x) dx
\label{eq:j3}
\end{equation}

$4 \pi J_3$ represents the mean number of excess galaxies around a
galaxy within a radius $r$.

Another important integral quantity is the volume--averaged two--point 
correlation function $\bar{\xi}(r)$:

\begin{equation}
\bar{\xi}(r) = \frac{1}{V^2} \int_V \xi(r_1 r_2) d^3r_1 d^3r_2
\label{eq:xibar}
\end{equation}

Contrarily to $\xi(r)$, 
$\bar{\xi}(r)$ is usually computed through counts in randomly placed
spherical cells, taking into account only the spheres totally included 
within the sample boundaries. No random catalog is used,
analogously to the estimate of $\Gamma(r)$.

The equivalent of $J_3(r)$ and $\bar{\xi}(r)$ in the fractal approach
is the integrated conditional density $\Gamma^*(r)$ defined as:

\begin{equation}
\Gamma^*(r) = \frac{3}{4 \pi r^3} \int _0 ^r 4 \pi x^2 \Gamma(x) dx
\label{eq:gammastar}
\end{equation}

which ``gives the behavior of the average density of a sphere of radius $r$
centered around an occupied point averaged over all occupied points"
(Coleman \& Pietronero 1992).

We notice that $\Gamma^*(r)$ is related to $J_3(r)$: using relations 
(\ref{eq:xigam}), (\ref{eq:j3}) and (\ref{eq:gammastar}) we have:

\begin{equation}
\Gamma^*(r) = n [1 + \frac{3}{r^3} J_3(r)]
\label{eq:gstarj3}
\end{equation}

which represents the integral counterpart of equation (\ref{eq:xigam}).

As in the case of $\xi(r)$, $J_3(r)$ is also computed using random samples,
i.e. assuming homogeneity beyond the sample limits, so
equation (\ref{eq:gstarj3}) cannot be used to obtain $\Gamma^*(r)$
from $J_3$.

Assuming now that $\xi(r)$ is a perfect power law,
we have the following relations:

\begin{equation}
J_3 (r) =  \frac {r^3}{3 (1-\gamma/3)} \xi(r)
\label{eq:j3xi}
\end{equation}

and

\begin{equation}
\bar{\xi} (r) =  
\frac {2^{-\gamma}}{(1-\gamma/3)(1-\gamma/4)(1-\gamma/6)} \xi(r)
\label{eq:xibarxi}
\end{equation}

We have then:

\begin{equation}
 \Gamma^*(r) = n [ 1 + 2^\gamma (1-\gamma/4)(1-\gamma/6) \bar{\xi}(r)]
 = n [1 + \frac{\xi(r)}{(1-\gamma/3)}]
\label{eq:gstarximed}
\end{equation}

Equations (\ref{eq:j3xi}--\ref{eq:gstarximed}) are useful
because they allow us to estimate $\xi(r)$ from the integral
indicators; in the next section we will use them,
with equation (\ref{eq:xigam}), to compare the different
functions we have defined above.

%
%
\section{Fractal analysis of the SSRS2}

\begin{figure}
\resizebox{\hsize}{!}{\includegraphics{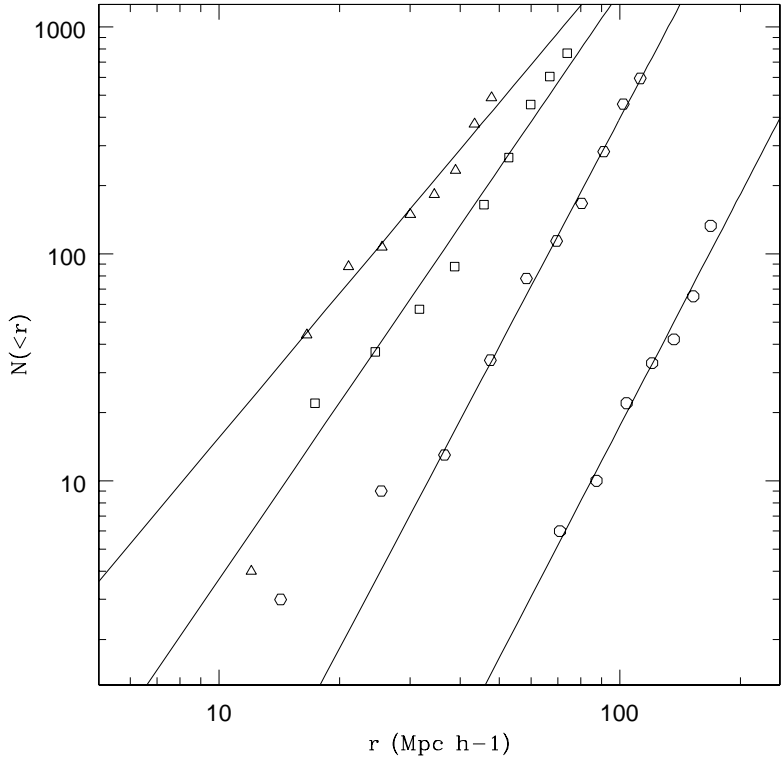}}
\caption[]{Integrated number of galaxies as a function of comoving
distance $r$ for 4 volume--limited subsamples of the SSRS2 ($M \le -18$,
$M \le -19$,$M \le -20$,$M \le -21$). The solid lines
are least--squares fits to the data with the law 
$N(<r) \propto r^{D}$, with respectively $D=2.11 \pm 0.27$, 
$D=2.59 \pm 0.21$, $D=3.34 \pm 0.18$, $D=3.40 \pm 0.35$.}
\label{fig:fracdens}
\end{figure}

A preliminary test --and the easiest one-- in order to compute the fractal
dimension of the galaxy distribution can be performed counting the number 
$N(r)$ of objects within a distance $r$ from us: 
from equation (\ref{eq:fracdens}),
we expect $N(r) \propto r^D$. The results for the 4 volume--limited
SSRS2 samples with absolute magnitude limits
$M \le -18$, $M \le -19$, $M \le -20$, $M \le -21$,
are shown in figure \ref{fig:fracdens}.
%
%
A formal least--squares fit to the data gives 
$D=2.11 \pm 0.27$, $D=2.59 \pm 0.21$, $D=3.34 \pm 0.18$, $D=3.40 \pm 0.35$;
the errors were computed through bootstrap resampling of the original samples.
These results apparently tell us that from the shallowest 
to the deepest sample we have a change in the fractal
dimension, with convergence to homogeneity.
Even considering the uncertainty associated to these values, 
we can conservatively state that
such results are more consistent with
homogeneity at scales $\sim 100$ \h,
than with a fractal Universe.

We have also computed $\Gamma(s)$ and $\Gamma^*(s)$ for 
the same 4 volume--limited samples. In what follows we will use
redshift space estimates.
First of all, we are interested here in a relative comparison
between different indicators, and not in determining a true value;
second, we want to compare our results with those of
Pietronero and collaborators, always performed in redshift space;
finally, the SSRS2 is not affected by large peculiar motions
(see Willmer et al. 1998).

%
%
\begin{figure*}
\resizebox{\hsize}{!}{\includegraphics{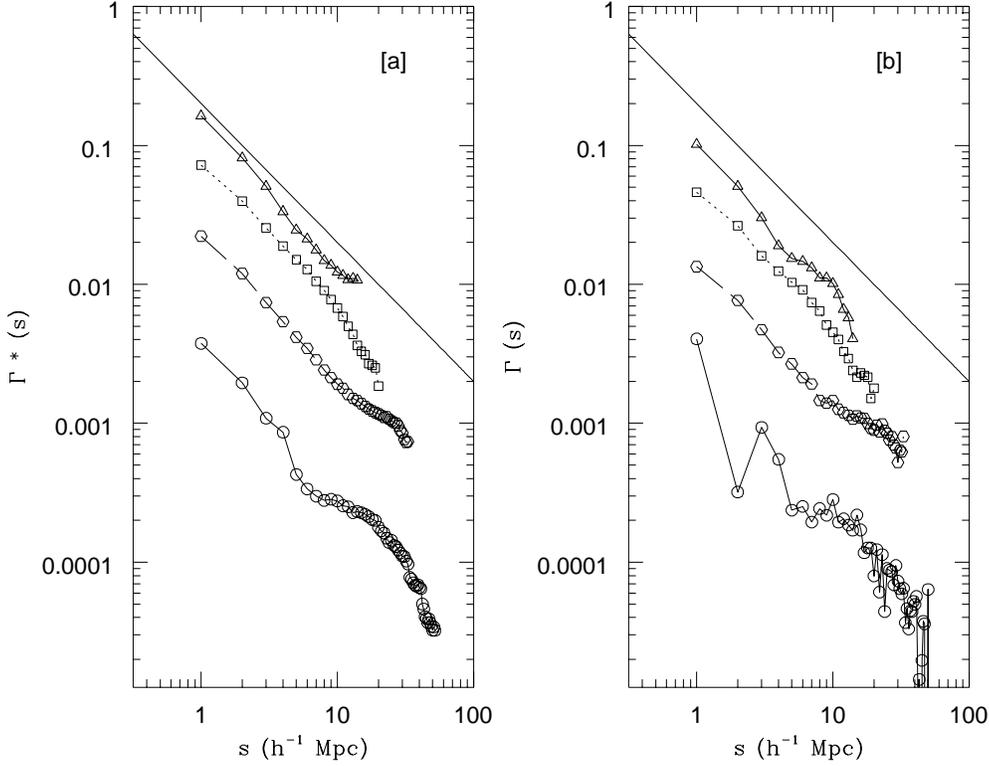}}
\caption[]{$\Gamma^*(s)$ (panel a) and $\Gamma(s)$ (panel b) for
the SSRS2 volume--limited samples including galaxies 
with $M \le -18$ (triangles), $M \le -19$ (squares), $M \le -20$
(hexagons), $M \le -21$ (circles). The solid line has
a slope $\gamma =-1$ (i.e. $D=2$).}
\label{fig:gamma}
\end{figure*}

The results are shown in figure \ref{fig:gamma}.
A least--square fit in the small--scale
range (less than 10 \h) gives a slope $\gamma = 3 - D \sim 1$, i.e.
$D \sim 2$. At larger scales,
the slope fluctuates from sample to sample,
and within the same sample, but does not show a clear {\em plateau}
(as should be expected if homogeneity was reached);
moreover, positive values of $\Gamma(s)$ and $\Gamma^*(s)$ 
are measured to larger scales for deeper samples.
These results are consistent with those found by Pietronero
and collaborators on other catalogs.

Does all this mean that the Universe is fractal, 
and that the hypothesis of large--scale homogeneity is falsified?

First of all, we have to take into account the behavior of
$N(r)$ we have discussed above, which is consistent with homogeneity
at large scales. $\Gamma(s)$ and $\Gamma^*(s)$ are more general tests,
as they average on all the galaxies, while $N(r)$ is centered on our
position; nevertheless, with $N(r)$ we can test larger scales ($ >
100$~\h), while $\Gamma(s)$ and $\Gamma^*(s)$ are limited to
$\sim 40$ \h. At this scale we still have a positive correlation function
(see below) and we should not expect to measure $D = 3$; therefore
what we can infer from all these tests is that there is still power at 
a scale of $\sim 40$ \h, but no evidence for significant power at 
$\sim 100$ \h.

A further problem for the fractal model is the discrepancy 
between the slope of $\xi(s)$, which implies $ D \sim 1$ (for the
SSRS2 see our results in table 1 of Paper I), and that
of $\Gamma(s)$ and $\Gamma^*(s)$, implying $D \sim 2$.
 SMP claim that people fit $\xi(s)$ in the ``wrong range" around $r_0$,
where $\xi(s)$ would deviate from a power--law. We notice that
people usually fit $\xi(s)$ where
it can be reasonably approximated by a power--law, excluding both small and
large scales; therefore the explanation of SMP (see their 
figure 9) does not seem plausible.

\begin{figure}
\resizebox{\hsize}{!}{\includegraphics{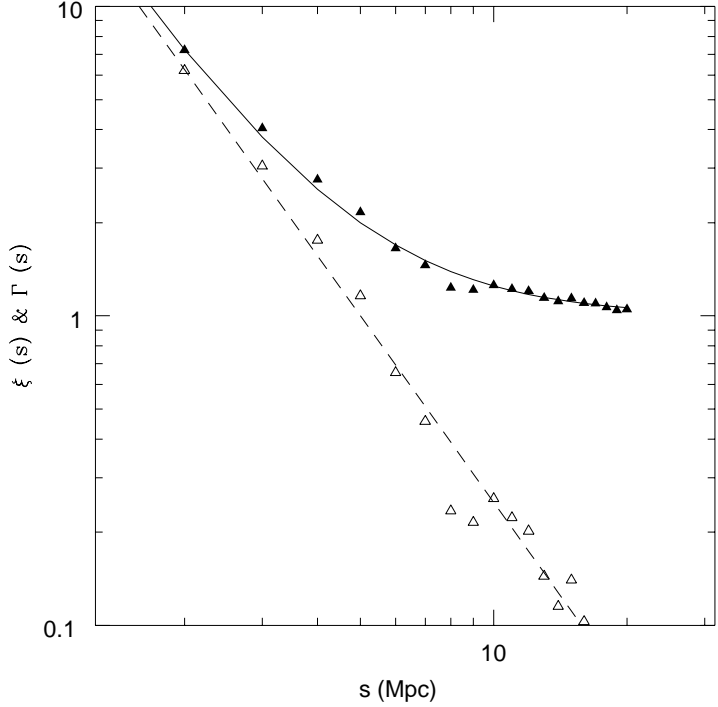}}
\caption[]{$\xi(r)$ (dashed line) and $\Gamma(r) = n[1+\xi(r)]$ (solid line)
assuming $\xi(r) = (r_0/r)^2$, with $r_0 = 5$\h and $n=1$.
Open and filled triangles simulate a measure of the above two
functions assuming a 20\% error.}
\label{fig:gammasim}
\end{figure}

In interpreting these results, it should perhaps be taken into account an
effect which is trivial but which we point out for sake of clarity.
In figure \ref{fig:gammasim}, we show $\xi(r)$ and $\Gamma(r)$ assuming
that $\xi(s)$ is a perfect power--law with $\gamma = -2$, and
points are obtained assuming an error of 20\%. It is obvious that
the quantity $\Gamma(r) \propto [1 + \xi(r)]$ can be approximated by 
2 power--laws; in this case, in the range 2--10 \h~ it has 
a slope $\gamma = -1.3$ or $D = 1.7$.
At larger scales, where $\xi(r)$ becomes very small, $\Gamma$ becomes constant.
However, in real samples $\xi(s)$ deviates from a power--law at
small and large scales (see below), and the regime where $\Gamma(r)$
is constant may not be reached.
%
%
This simple exercise tells us that if $\xi(s)$
is a genuine power--law, there is a range around
$r_0$ where $\Gamma(s)$ can be approximated by a 
power--law with a flatter slope, which is what is observed.

\begin{figure}
\resizebox{\hsize}{!}{\includegraphics{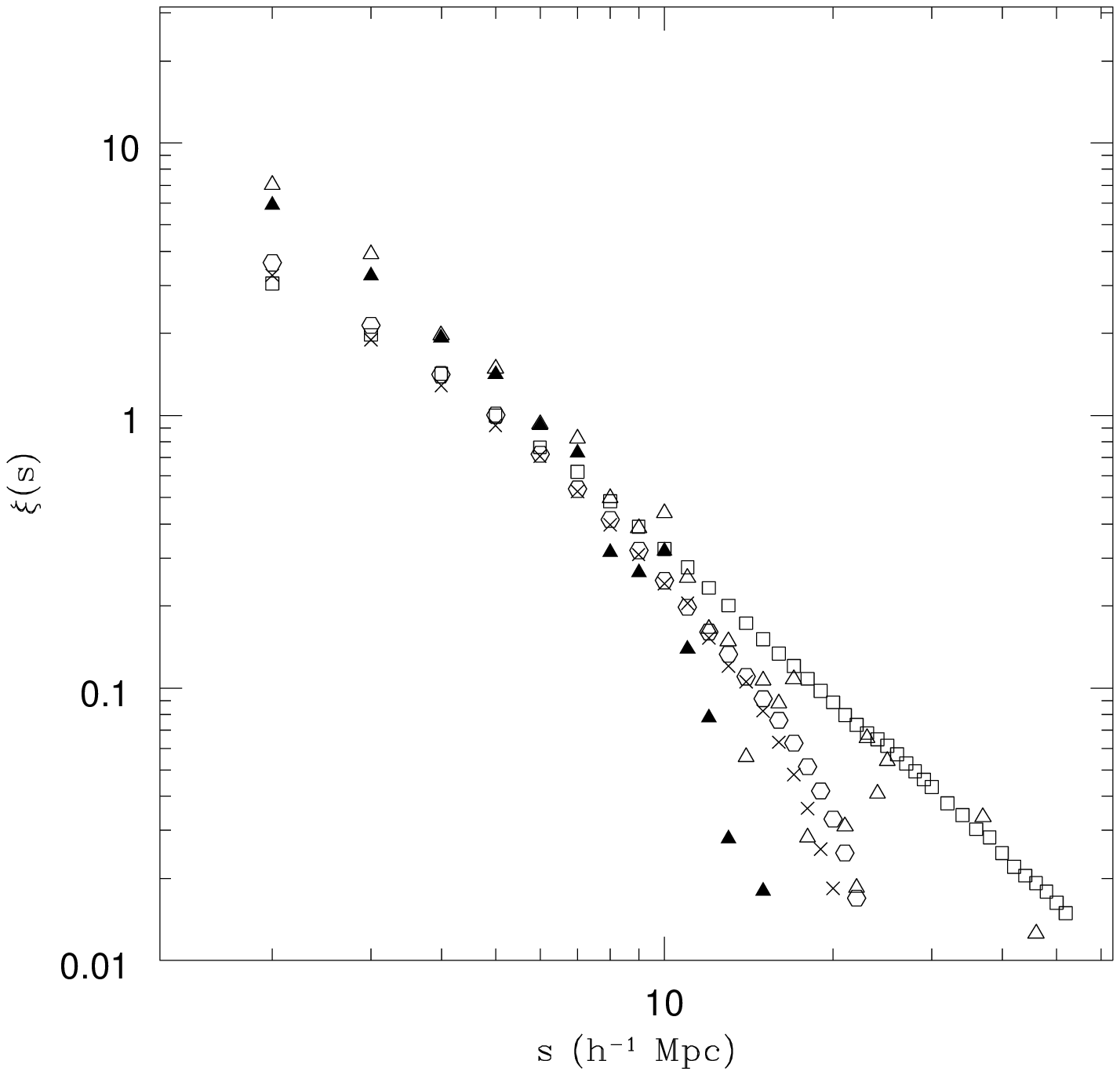}}
\caption[]{$\xi(s)$ derived with different statistical tools for
the $M=-20$ subsample of the SSRS2.
Open triangles: direct standard estimate; 
filled triangles: computed from $\Gamma$; open squares: computed from $J_3$; 
open hexagons: computed from $\bar{\xi}$; crosses: computed from $\Gamma^*$.}
\label{fig:xiall}
\end{figure}

We will check now the internal consistency between standard
and non standard statistics, examining the behavior of the
functions described in the previous section.
In figure \ref{fig:xiall}, for the
volume --limited subsample including all galaxies with $M \le -20$,
we show the values of $\xi(s)$ estimated
directly using the Hamilton (1994) formula,
from $\Gamma(s)$ through equation (\ref{eq:xigam}), 
and from the integral functions $J_3(s)$, $\bar{\xi}(s)$ and
$\Gamma^*(s)$ using equations  (\ref{eq:j3xi}), (\ref{eq:xibarxi}),
(\ref{eq:gstarximed}) (i.e., assuming that $\xi(s)$ is a perfect power
law).
In the standard scenario,
all these estimates applied on the same sample should in principle measure 
the same quantity, and the only differences should be due to the biases 
introduced by each method (this is why no error bars are shown in the
figure).

%
%

Figure \ref{fig:xiall} shows that there are two
main effects:
at small scales, there is a systematic difference between the
estimates of $\xi(s)$ obtained from the integrated functions and the
two estimates obtained directly or from $\Gamma(s)$; at large scales,
there is a systematic difference between the estimates based on
random catalogs and the other ones which only take into account the
volumes fully included within the sample limits. 

The direct estimate of $\xi(s)$ is consistent
with that computed from $\Gamma(s)$ to about $10$\h. 
At small scales peculiar motions affect $\xi(s)$, 
and the number of pairs at small separations becomes very small; 
therefore we expect significant deviations of $\xi(s)$ from a power--law, and
this explains why the indirect estimates of $\xi(s)$
from $\Gamma^*(s)$, $\bar{\xi}(s)$, and $J_3(s)$
based on the assumption of a perfect power--law, 
differ systematically from the other two estimates.

At scales larger than $\sim 10$\h~ the estimates of
$\xi(s)$ obtained from $\Gamma(s)$,
$\Gamma^*(s)$ and $\bar{\xi}(s)$ decline more rapidly than the direct estimate
of $\xi(s)$ and that obtained from $J_3(s)$.
As the former indicators take into account
only the volumes fully included within the sample
limits, they are limited to smaller scales 
--their maximum scale corresponds to the
radius of the largest sphere which can be included in the sample--; 
moreover, at large scales the randomly placed spheres have a significant
overlap, and are no more independent, thus giving a biased (lower) 
estimate of the variance. It can be noticed that the estimate of $\xi(s)$
from $\Gamma(s)$ drops somewhat more rapidly than the 
estimates from $\Gamma^*(s)$ and $\bar{\xi}(s)$, because 
the latter are derived from integral measures, 
thus are smoother and have a better signal to noise ratio. 
It is also clear that the estimate of $\xi(s)$ from
$J_3(s)$, which is an integrated measure and is based
on random catalogs, can reach the largest scales, and
appears to be consistent with a power--law extending to $\sim 50$ \h.
The direct standard estimate of $\xi(s)$
is noisier and oscillates; it drops beyond $10$ \h, in a way similar to
the integral estimates, then has a second oscillation, following
the $J_3(s)$ estimate.

We can conclude from the above discussion that there is no disagreement
between the results obtained with standard and non--standard indicators,
when taking into account the range of validity of each one of them.

\begin{figure}
\resizebox{\hsize}{!}{\includegraphics{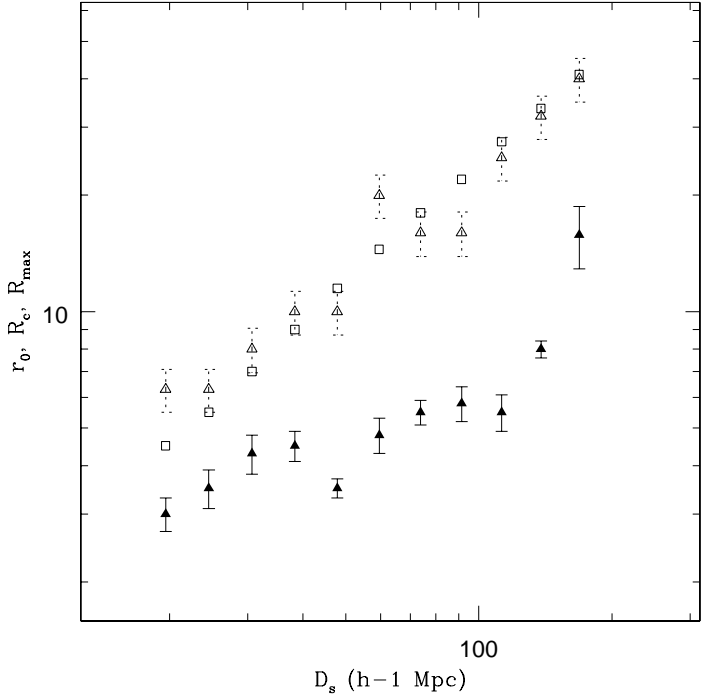}}
\caption[]{The correlation length $r_0$ (filled triangles), 
the zero--point $R_c$ (open triangles), and the maximum radius 
$R_{max}$ (open squares) as a function of the sample depth $D$, for
volume--limited samples of the SSRS2 (from $M \le -16$ to $M \le -21$,
with a step of $0.5$ magnitudes).}
\label{fig:rc}
\end{figure}

In figure \ref{fig:rc}, for the volume--limited subsamples
from $M=-16$ to $M=-21$ (step 0.5) we show 
the correlation length $r_0$, the first zero--point $R_c$ of $\xi(s)$
(i.e. the scale where $\xi(s)$ becomes negative),
and the radius of the largest sphere totally included in the sample, 
$R_{max}$, as a function of the sample depth $D_s$.
The values of $r_0$ with their bootstrap error bars are the same as in
figure \ref{fig:depth}, while the values of $R_c$ simply
correspond to the last bin where $\xi(s)$ is positive, and their
error bars indicate the bin width.

%
%
It is trivial that $R_{max} \propto D_s$. The correlation length $r_0$
is not proportional to the depth, contrarily to what expected for a 
fractal; the less deep subsamples are probably not ``fair'', and $r_0$ may be
underestimated (even if a luminosity effect for low luminosity
galaxies cannot be excluded); then $r_0$ is more or less constant in
an intermediate range of luminosities, and increases rapidly (more
rapidly than the depth) for samples including galaxies brighter than $M_*$.
This is the luminosity effect we have shown in Paper I (see also figure
\ref{fig:depth}).
The most striking result of figure \ref{fig:rc} 
is that the value of the zero--point $R_c$ clearly 
increases linearly with the depth, 
and that its value corresponds in a striking way to $R_{max}$.

This quantitative correspondence between $R_c$ and $R_{max}$
means that the standard estimate of 
the correlation function $\xi(s)$
is in fact reliable only at scales smaller than the largest sphere contained 
in the sample: it is at this scale which $\xi(s)$ 
is forced to become negative, then starts to oscillate.
We stress that this effect should be taken into account; often 
an oscillating correlation function is interpreted as evidence of
a real characteristic scale, but if this scale
corresponds to the $R_{max}$ of the sample taken into account, the
unavoidable conclusion (as in the case of our subsamples) is that
the real zero--point of $\xi(s)$ has not been reached.

%
%
\section{Comparison with theoretical predictions}

Are the results of our fractal analysis consistent 
with standard models of galaxy formation?

\begin{figure}
\resizebox{\hsize}{!}{\includegraphics{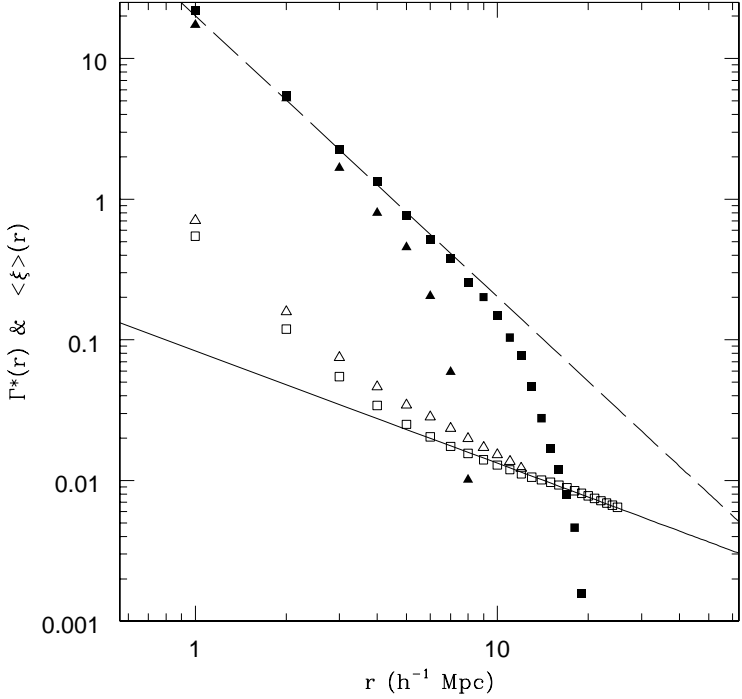}}
\caption[]{$\Gamma^*(r)$ and $\xibar(r)$ measures for one realization of a CDM
universe. Filled and open triangles represent
respectively $\xibar$ and $\Gamma^*$ measured in a box with $L =25$ \h;
filled and open squares represent respectively $\xibar$ and $\Gamma^*$ 
measured in a box $L =50$ \h.
The solid line has a slope $\gamma = -0.8$; the dashed line has a 
slope $\gamma = -2.0$.}
\label{fig:fracsim}
\end{figure}

The behavior of $\Gamma$ and $\Gamma^*$
can be checked by using N--body simulations, which are based
on standard Friedmann--Lema\^{\i}tre cosmological models, and
assume homogeneity at large scales.
We used the Adaptive $P^3M$ code of Couchman (1991) to evolve
a standard Cold Dark Matter universe. We started from the initial
linear power spectrum of Bond \& Efstathiou (1984) with $\Omega h = 0.5$.
The box size was 100 \h, with $64^3$ particles.
We analyzed the simulation at a time when its correlation function
was roughly comparable to the observed one, diluting it
to approach the observed galaxy densities.
Figure \ref{fig:fracsim} shows $\xibar(r)$ and $\gammastar(r)$
for 2 subsamples extracted from the simulation:
the first with a box size $L = 25$ \h, the second
with $L = 50$ \h.
%
%
The best--fit slope for $\Gamma^*$ is $\gamma = -1.0$, i.e. $D = 2$ in
the first case; $\gamma = -0.8$, i.e. $D \sim 2.2$, in the second case.
$\xibar(r)$ shows the typical cutoff when approaching the radius of
the largest sphere which can be contained in the sample: in fact,
the deeper sample allows to recover power at larger scales.
These results are comparable to our results on the SSRS2, and also
to those presented by SMP.
We have already mentioned that a problem for a simple fractal Universe 
is the discrepancy between the fractal correlation dimension implied
by $\xi(s)$, $D = 3 - \gamma \sim 1$, and  that
derived from $\Gamma(s)$ or $\Gamma^*(s)$, $D \sim 2$. We
have also noted that such a discrepancy cannot be due simply to an 
improper fitting of $\xi(s)$  (as claimed by SMP)
in a range where it deviates from a power--law.

Now we have shown that the observed slopes of $\xi(s)$ and $\Gamma(s)$ can be
reproduced by a standard CDM N--body simulation: this confirms the
implications of our example in figure \ref{fig:gammasim}, 
and the consistency of both $\xi(s)$ and $\Gamma(s)$
with gravitational clustering in a Universe
homogeneous at large--scale.

We have not addressed, of course, 
the question of the variations of $r_0$. 
In our simulation, $r_0$ at any fixed time has a unique, constant value;
when analyzing deeper subsamples of our simulation, 
$r_0$ rapidly converges to its true value 
(no significant difference in fact is found in the value $r_0$ measured in a 
box of $50$ \h~ and in another at $75$ \h), while the scale of the
zero--point increases significantly.

As we have previously shown, in the real Universe 
the correlation length $r_0$ of galaxies does depend on their
luminosity (or, in other terms, there is evidence of luminosity segregation).
Our simple answer to the criticism by
SMP --``the authors [...] have never presented
{\em any quantitative argument that explains the shift of $r_0$ 
with sample size}''--,
is that shallow samples are affected by the
presence of local structures, as the discrepancy between the 
Perseus--Pisces and the other surveys reminds us; moreover,
when a sample is not deep enough, $\xi(s)$
can be artificially truncated, thus affecting the estimate of $r_0$;
finally, and it is our main quantitative argument,
from figure \ref{fig:depth} it is clear that $r_0$ does not depend
on sample depth, but on galaxy luminosity (as shown in paper I),
for $L > L_*$ galaxies.

On the other hand, there are various reasons why we should
expect luminosity segregation.
It has been shown that the correlation function can be
amplified as a consequence of a ``statistical bias" (Kaiser 1984), which
can explain in a natural way the cluster autocorrelation function.
As concerning galaxies, White et al. (1987) used N--body simulations
to show that more massive galaxies are
more clustered; they called this phenomenon ``natural bias''.
Recent semi--analytical models
(Mo \& White 1996, Kauffmann et al. 1997) predict this bias if
galaxies form in sufficiently massive halos, with $M > M_*$; in this case,
the trend is similar to that observed in the SSRS2  (see
figure 5 of Paper I), even if they do not succeed in fitting both the low and
high luminosity samples. This is not surprising, given all the
theoretical uncertainties, particularly
on the dark halo--galaxy relation.

In addition to the luminosity--dependent correlation amplitude,
there are other 2 important points which should be considered when
analyzing the galaxy distribution: 
the K--correction and the cosmological model (see Guzzo 1997;
Scaramella et al. 1998).
While in our analysis of the SSRS2 samples we have fixed an 
$H_0=100$, $q_0=0.5$ model and systematically applied K--corrections,
Pietronero and collaborators usually perform their analysis
in Euclidean space and without K--correction.
As concerning the cosmological model, for a fractal Universe
one should find and apply the appropriate formulae for an isotropic but 
inhomogeneous Universe (unless one wants to question the validity of General 
Relativity).
On the other hand, the K--correction is an empirical correction
to the magnitude of a galaxy, as in a fixed wavelength range
we observe different parts of the spectrum at different
redshifts.
The effect is still small in the
deepest sample of the SSRS2; if we use Euclidean distances and do not
apply the K--correction, from the relation $N(<r) \propto r^D$
we find $D = 3.28$ instead of $D = 3.40$.
It becomes critical in deeper surveys as the 
ESO Slice Project (Vettolani et al. 1997; Zucca et al. 1997), as shown by 
Scaramella et al. (1998), who do not
confirm the results obtained by SMP on that sample
as well as on Abell clusters.

Moreover, the K--correction cannot be ignored when interpreting 
deep counts; the claim by Sylos Labini et al. (1996) that deep counts
are consistent with a fractal distribution can be reversed, 
concluding that the agreement with the observed count slope would be lost 
if at least the K--correction had been applied (not to speak of
cosmological and evolutionary corrections: they  are model--dependent, but we
know that galaxies evolve).

Why do the choice of an Euclidean space and the absence of a
K--correction give results more consistent with a fractal distribution?
There are essentially two main reasons: first of all,
Euclidean distances become systematically larger than comoving
ones at increasing redshift: this means also that one will have larger
voids and structures. 
Second, when applying the K--correction we can recover
galaxies that otherwise would not have been included in a 
volume--limited subsample; if the K--correction is not applied,
a systematically larger fraction of galaxies will be lost at larger
distances. The combination of the two effects lowers the density
at increasing redshift, and flattens the slope of the 
$\log(N)-\log(r)$ relation, thus
following the same trend of a fractal distribution.

%
%
\section{Conclusions}

In this paper, we have carefully examined the claim that 
the Universe is a fractal, 
analyzing the SSRS2 and using the same statistical approach of
Pietronero and collaborators.

Here are our main results:

\begin{itemize}

\item the SSRS2 has a luminosity dependent correlation amplitude;

\item results obtained with the standard and fractal approach
      give consistent results;

\item concerning large--scale homogeneity,
      observational evidence is --for the moment--
      consistent with theoretical predictions
      derived from standard CDM models of galaxy formation,
      and does not require a fractal Universe;

\item on the other hand, the first zero--point of the correlation
      function we measure in SSRS2 volume--limited subsamples
      increases linearly with the sample depth, and
      has approximately the same value of the radius of the largest sphere 
      which can be included in the sample.
      This implies that the zero--point of $\xi(s)$ is beyond 
      40 \h, a lower limit set by very luminous galaxies
      (see Benoist et al. 1996; Cappi et al. 1998) and
      consistent with results from clusters 
      (e.g. Cappi \& Maurogordato 1992);
      on this point, we agree with SMP that measurements of $\xi(s)$ beyond a 
      scale corresponding to the radius of the largest sphere included in
      the sample are not reliable.

\end{itemize}

We conclude that the dependence of the galaxy correlation
function on the galaxy luminosity and the power still present at large scales 
($\ge 40$ \h) can satisfactorily explain those effects interpreted by SMP as 
evidence of an inhomogeneous, fractal Universe.

\begin{acknowledgements}
We thank F. Sylos Labini and L. Pietronero for many useful discussions.
AC would like to thank the hospitality of the Observatoire de Nice.
\end{acknowledgements}

\end{document}